\documentclass[final,5p,times]{elsarticle}
\usepackage{graphicx}
\usepackage{amssymb}
\usepackage{verbatim}
\usepackage{multirow}
\usepackage{tabularx, colortbl}
\usepackage[hidelinks]{hyperref}
\usepackage[all]{hypcap}

\def\svu{cm$^{3}$~s$^{-1}$}
\def\sv{$\langle \sigma v \rangle$}

\begin{document}
\begin{frontmatter}

\title{A decade of dark matter searches with ground-based Cherenkov
  telescopes\tnoteref{title}}
\tnotetext[title]{The original title of the invited talk was "Dark matter
  searches with Cherenkov Telescopes". This version has some typo
  corrections with respect to the published version, as well as an
  updated Fig.~2.}

\author{Michele Doro}

\address{University and INFN Padova, via Marzolo 8, 35131 Padova (Italy)\\
Department of Physics and CERES, Campus Universitat Aut\'onoma
Barcelona, 08135 Bellaterra (Spain)}

\begin{abstract}
In the general scenario of Weakly Interacting Massive Particles
(WIMP), dark matter (DM) can be observed via astrophysical gamma-rays
because photons are produced in various DM annihilation or decay
processes, either as broad-band or line emission, or because of the
secondary processes of charged particles in the final stages of the
annihilations or the decays. The energy range of the former processes
is accessible by current ground-based Imaging Atmospheric Cherenkov
telescopes (IACTs, like H.E.S.S., MAGIC and VERITAS). The strengths of
this technique are: a) the expected DM gamma-ray spectra show peculiar
features like bumps, spikes and cutoff that make them clearly
distinguishable from the smoother astrophysical spectra, b) the
expected DM spectrum is universal and therefore by observing two or
more DM targets with the same spectrum, a clear identification
(besides detection) of DM would be enabled. The role of IACTs may gain
more importance in the future as the results from the LHC may hint to
a DM particle with mass at the TeV or above, where the IACTs sensitivity
is unsurpassed by other experiments. In this contribution, a review of
the search for DM with the current generation of IACT will be
presented. 
\end{abstract}

\begin{keyword}
IACT \sep gamma-ray astronomy \sep dark matter \sep H.E.S.S. \sep
MAGIC \sep VERITAS \sep CTA


\end{keyword}

\end{frontmatter}


\section{Introduction}
\label{sec:intro}
\vspace{-3mm}
One of the most interesting and compelling observations that
ground-based Cherenkov telescopes operating in the very-high-energy
gamma-ray band can perform is that looking at targets in the sky
where a large concentration of dark matter (DM) is expected. There are
several reasons in support of 
this. {\bf First} of all, DM is indeed expected in the sky. 
80\% of the total
matter content of the Universe is constituted by one or more new types of
particles. The DM has shaped the formation of the first stars
and galaxies, so thoroughly that the concordance cosmological model is
called $\Lambda$CDM where CDM stands for Cold DM. 
We also know that there are places in the sky where DM is expected to
be particularly concentrated. We don't know the DM nature and
if it could finally be detected via primary or secondary radiation 
associated with its annihilation or decay, but there are several models
that predict such signatures and it is therefore worth ``sailing'' our
telescopes to these promised lands, despite no ``Earth!'' signal has
arrived so far.
{\bf
  Secondly}, the gamma-ray band is a very privileged one, for 
several reasons: a) gamma-rays are neutral and trace back to the point
of origin, where we expect DM, b) the gamma-ray spectrum
emerging from DM interactions
(either annihilations or decays) is
universal. All DM targets are expected to show exactly the same
gamma-ray spectrum. The observation of multiple spectra from different
targets would therefore constitute an excellent result, c) gamma-ray
spectra from DM annihilations or decay typically 
show several characteristic features, naturally depending on the specific dark
matter type, but in general classifiable in sharp cutoff, bumps, or
even line emissions. This makes the DM spectra hardly
confusable with typical astrophysical spectra. {\bf Third}, the recent
experimental results of the LHC experiments: the quite large Higgs
boson mass and the non-evidence for New Physics beyond the Standard
Model are possibly hinting to DM particle being more massive
than expected, about the TeV or above~\cite{Feng:2013pwa}. The TeV region is where ground-based
telescopes have highest sensitivity. 

And indeed this is what was done in the last decade, specially with
the H.E.S.S., MAGIC and VERITAS experiments. These very successful
experiments, all together, invested quite a large fraction of their
observation times in the last years to cover 
the targets where DM was expected.
In this contribution, we will try to review these
observations. For additional details on gamma-ray signals from dark
matter, we refer the reader to Ref.~\cite{Bringmann:2012ez}.


\vspace{-2mm}
\section{Ground-based imaging Cherenkov telescopes}
\label{sec:iact}
\vspace{-3mm}
Gamma rays produced in Space cannot cross the Earth atmosphere. After few radiation lengths,
they interact with the electrostatic field of the atmospheric atoms
and convert into an electron-positron pair. However, these leptons are
extremely energetic (they share the energy of the primary
gamma ray), and therefore are able to initiate an electromagnetic
showers thanks to subsequent emission of bremsstrahlung gamma rays
which in turn pair produce again. The shower is few km long and few
hundreds of meter large. It has a maximum at about 10-12 km above the ground
(at the GeV--TeV) and dies out when the lepton energy reaches the
threshold for ionization, which in air is 83 MeV. Even if the shower
particles are lost in the atmosphere, the fact that most of the time
the leptons were travelling faster than the speed of light in the
atmosphere, allowed for the radiation of Cherenkov light from the
atmospheric medium swept up by the shower. This light has a continuous
spectrum, peaked at about 300~nm after the ozone absorption, and it is
a flash of 
light of the duration of few ns, expanding within a cone of aperture
of about 1~deg radius, which at the ground illuminates an area of
about 100~m radius. For this reason, if one puts a telescope in this
Cherenkov light pool, the ``effective'' area of the telescope is not
simply the geometrical area of the dish, but almost the entire
Cherenkov pool, which sums up to about $10^5$~m$^2$. One could compare
this number with the typical areas of satellite-borne gamma-ray
detectors, of about $1-2$~m$^2$, hardly expandable in the near
future. The drawback is that gamma-ray showers at few GeV produce too
little Cherenkov photons to be detected, and the technique is fully
sensitive at the TeV scale and above.


Ground-based Cherenkov telescopes are also called Imaging Atmospheric
Cherenkov Telescopes (IACTs) because they ``image'' the shower, in the
sense that if a multi-pixel camera is placed at the telescope focal
plane, the image of a typical electromagnetic shower is an oblate ellipse
pointing to the center of the camera (corresponding to an inclined
section of the Cherenkov light cone). Through the careful analysis of
shower primary, secondary and sometimes tertiary moments, and in some
cases its time evolution, one can reconstruct the energy and direction of the
primary gamma ray. 
The main source of background is constituted by the showers generated
by hadronic cosmic rays. These are mixed sub-hadronic and sub-electromagnetic showers
initiated by primary cosmic rays in the top atmosphere like protons,
helium, and heavier nuclei. They are by far more abundant than
gamma rays, which are seen roughly every 10,000 hadrons. Therefore,
the technique relies on a first telescope topological trigger system
that rejects about 99\% of the hadrons, and later on a gamma/hadron
separation at the analysis level.
The technique was pioneered by the Whipple observatory, that after
tens of years of hunting, finally detected the Crab Nebula in 1989, and
it is now superseded by the H.E.S.S., MAGIC and VERITAS arrays.

Table~1 collects some information regarding the major IACT experiments. We mention also that the Whipple telescope terminated operation this year. 
\begin{table}[h!t]
\centering
\small{%
\begin{tabular}{l|ccc}
\hline
IACT    & Year & Nr. tels \& diameter & Location \\
\hline
Whipple & 1968 & 1$\times$12~m & Arizona, USA \\
H.E.S.S.    & 2003 & 4$\times$12~m+1$\times$28 m & Gambserg, Namibia \\
MAGIC   & 2004 & 2$\times$17~m   & La Palma, Spain \\
VERITAS & 2007 & 4$\times$12~m   & Arizona, USA \\
\hline 
\end{tabular}
}
\vspace{-2mm}
\caption{\label{tab:iact}Current major operating ground-based
  Cherenkov telescopes. Given are the starting year, the array
  multiplicity and dish diameter \emph{in the latest configuration},
  and the location.}
\end{table}

\section{Gamma-ray signatures from dark matter}
\label{sec:gammaray}
\vspace{-3mm}
There is no space in this review to account for all the particles
that are valid candidate for DM. We refer to
\cite{Feng:2010gw} for a recent review. In this context, we
concentrated more on a Weakly Interacting Massive Particle (WIMP)
scenario, which foresees a particle at the GeV-TeV
scale, whose annihilation or decay products are found in the Standard Model
(SM) particle zoo. There are valid scenarios of decaying DM, however,
they are not furtherly treated here.

\begin{figure}[h!t]
\centering
\includegraphics[width=0.9\linewidth]{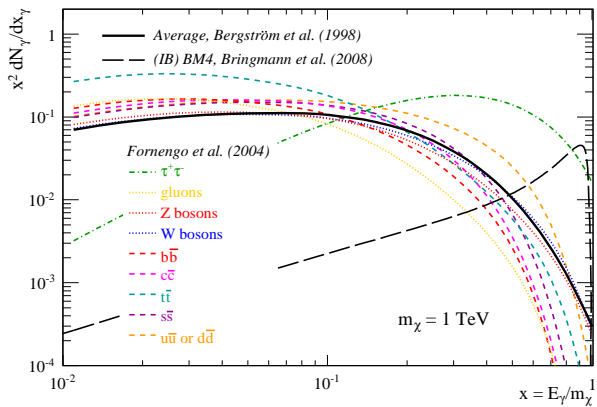}
\vspace{-3mm}
\caption{\label{fig:spectra} Differential spectra (multiplied by $x^2$)
  of gamma-rays from several neutralino annihilation
  products. Taken from~\cite[Fig.~1]{Charbonnier:2011ft}.} 
\end{figure}

Now, if DM is coupled to the SM with some interactions, in the final
products of annihilations or decays one can find either leptons or
hadrons or gauge bosons, often the heaviest one because of the scale
of the DM mass, which is at the GeV-TeV. It is therefore expected
naively to find also gamma-rays in the final products. More precisely,
the gamma-ray emission can be originated as follows (see
Fig.~\ref{fig:spectra}): $a)$ from neutral pion
decays after hadronization of quarks. This   gives origin to a
broadband spectrum terminating with a cutoff at   the DM mass; $b)$
from final state radiation of leptons. This also
  gives origin to a broadband spectrum and a cutoff but with harder
  photons; 
$c)$ gamma rays from internal bremsstrahlung, when the annihilation
  is to sfermions and the annihilation is in the t-channel, which
  gives rise to a pronounced bump of gamma rays toward the mass cutoff~\cite{Bringmann:2012ez};
$d)$ from line-processes ending in $\gamma X$,
  where $X$ could be $\gamma, Z_0, h$. These are loop processes, whose
  intensity strongly depends on the specific DM realization, which
  give rise to line emissions that constitute smoking guns detection
  for DM, because an astrophysical explanation of these lines
  would be extremely challenging.
There are other gamma-ray emission processes, following other particle
models, which are not described here.

All in all, IACTs observe photons. Therefore, every process is valuable
as long as it provides enough photons to detect. However, the more
features the spectrum exhibits, the easier is the consequent DM
identification. In this sense, not only the total flux is relevant.
Generally, the DM annihilation flux in gamma rays is
expressed as:
\begin{equation}\label{eq:dmflux1}
\frac{d\Phi}{dE}(E;\Delta\Omega) = \frac{B_F}{4\pi} \frac{\langle\sigma_{ann}                  
  v\rangle}{2\, m^2_\chi}\frac{dN_\gamma}{dE} \int_{\Delta\Omega} \int_{los}d\theta 
ds\,\rho^2(\theta,s)
\end{equation}
where $\langle\sigma_{\rm ann} v\rangle$ is the averaged annihilation
cross-section times the velocity, $m_\chi$ the DM particle
mass, $dN_\gamma/dE$ is the total number of photons
  produced during one annihilation event, and their product is
  conventionally called the particle physics factor. The second term
  of the equation is called the \emph{astrophysical factor} or
  \emph{J-factor} and it is computed as the line of sight $s$
integral of the \emph{square} of the DM density, and over a certain
solid angle $\Delta\Omega=2\pi(1-\cos(\theta))$ under which the source is observed. 
Finally, $B_F$ is the so-called \emph{intrinsic boost factor}, and it
is a measure of the uncertainty in either the particle physics or the
astrophysics terms for unaccounted intrinsic contributions to the flux.
In case of of {\bf decaying}  DM, in the particle physics
factor of Eq.~\ref{eq:dmflux1}, the term $\langle\sigma_{\rm ann} v\rangle/m^2_\chi$ is
replaced by $\Gamma_{\rm dec}/m_\chi$ where $\Gamma_{\rm dec}$ is the
inverse of the particle lifetime, and the dependence on the DM
density $\rho$ is \emph{linear} and not quadratic.

\section{Overview of current experimental results}
\label{sec:overview}
\vspace{-3mm}
The search for DM with Cherenkov telescopes has started
together with the full establishment of the IACT technique, and namely
with the H.E.S.S., MAGIC, VERITAS and Whipple
experiments. Table~\ref{tab:targets} lists the various targets and
searches performed, organized by target class, and reporting
the year and duration of the observation (in hours), and the reference
publication (or proceedings). I apologize for possible missing entries in this table.

As a general trend, the searches started with the target classes that
10 years ago were supposed to be the best candidate: the galactic
center, galaxy clusters and dwarf satellite galaxies. More recently,
the attention has been also focused on different kinds of searches,
like those at intermediate mass black holes or DM subhalos and
at signatures of line-emission. Finally, -- specially raised
by the all-electrons and positron ratio anomalies observed in the past
5 years, discussed later-- the DM searches have also been extended to
signatures from cosmic ray leptons rather than gamma-rays, an
observation which is also possible with IACTs. In the following, an
overview class by class is given. 

\begin{table}[!t]
\centering
\small{%
\begin{tabular}{lccll}
\hline
\rowcolor[gray]{.8} 
{\bf Target} & {\bf Year} & {\bf Time} & {\bf Experiment} &
{\bf Ref.} \\
\hline
\multicolumn{5}{>{\columncolor[gray]{.92}}c}{{\bf Globular Clusters}} \\
\hline
M15 & 2002 & 0.2 & Whipple & \cite{Wood:2008hx} \\
    & $2006-2007$ & 15.2 & H.E.S.S. & \cite{Abramowski:2011hh} \\
M33 & $2002-2004$ & 7.9 & Whipple & \cite{Wood:2008hx} \\
M32 & 2004 & 6.9 & Whipple & \cite{Wood:2008hx} \\
NGC 6388 & $2008-2009$ & 27.2& H.E.S.S. & \cite{Abramowski:2011hh} \\
\hline
\multicolumn{5}{>{\columncolor[gray]{.92}}c}{{\bf Dwarf Satellite Galaxies}} \\
\hline
Draco & 2003 & 7.4 & Whipple & \cite{Wood:2008hx}\\
      & 2007 & 7.8&  MAGIC & \cite{Albert:2007xg}\\
      & 2007 & 18.4 & VERITAS & \cite{Acciari:2010ab}\\
Ursa Minor & 2003 & 7.9 & Whipple & \cite{Wood:2008hx}\\
 & 2007 & 18.9 & VERITAS & \cite{Acciari:2010ab}\\
Sagittarius & 2006 & 11 & H.E.S.S. & \cite{Aharonian:2007km}\\
Canis Major & 2006 & 9.6 & H.E.S.S.  & \cite{Aharonian:2008dm}\\
Willman 1 & $2007-2008$ & 13.7 &VERITAS & \cite{Acciari:2010ab}\\
 & 2008 & 15.5 & MAGIC  & \cite{Aliu:2008ny}\\
Sculptor & 2008        & 11.8 & H.E.S.S. & \cite{Abramowski:2010aa}\\
Carina   & $2008-2009$ & 14.8 & H.E.S.S. & \cite{Abramowski:2010aa}\\
Segue 1  & $2008-2009$ & 29.4 & MAGIC & \cite{Aleksic:2011jx}\\
         & $2010-2011$ & 48 & VERITAS & \cite{Aliu:2012ga}\\
         & $2010-2013$ & 158 & MAGIC & \cite{Aleksic:2013segue}\\
Bo\"otes & 2009 & 14.3 & VERITAS & \cite{Acciari:2010ab}\\
\hline
\multicolumn{5}{>{\columncolor[gray]{.92}}c}{{\bf Galaxy Clusters}} \\
\hline
Abell 2029 & $2003-2004$ & 6     & Whipple & \cite{Perkins:2006fa}\\
Perseus & $2004-2005$    & 13.5 & Whipple & \cite{Perkins:2006fa}\\
        & $2008$         & 24.4 & MAGIC   & \cite{Aleksic:2009ir}\\
Fornax  & $2005$         & 14.5 & H.E.S.S.    & \cite{Abramowski:2012au}\\ 
Coma    & $2008$         & 18.6 & VERITAS & \cite{Arlen:2012} \\
\hline
\multicolumn{5}{>{\columncolor[gray]{.92}}c}{{\bf The Milky Way
    central region}} \\
\hline
MW Center      & 2004        & 48.7 & H.E.S.S. & \cite{Aharonian:2006wh} \\
MW Center Halo & $2004-2008$ & 112  & H.E.S.S. & \cite{Abramowski:2011hc}\\
\hline
\multicolumn{5}{>{\columncolor[gray]{.92}}c}{{\bf Other searches}} \\
\hline
IMBH & $2004-2007$ & 400 & H.E.S.S. & \cite{Aharonian:2008wt}\\
     & $2006-2007$ & 25 & MAGIC & \cite{Doro:2007icrc}\\
Lines & $2004-2008$ & 112 & H.E.S.S. & \cite{Abramowski:2013ax}\\
      & $2010-2013$ & 158 & MAGIC & \cite{Aleksic:2013segue}\\
UFOs & -- & -- & MAGIC   & \cite{Nieto:2011uj} \\
     & -- & -- & VERITAS & \cite{GeringerSameth:2013fra}\\
All-electron & $2004-2007$ & 239 & H.E.S.S. & \cite{Aharonian:2009a,Aharonian:2008aa}\\
             & $2009-2010$ & 14 & MAGIC & \cite{BorlaTridon:2011dk}\\
Moon-shadow & -- & -- & MAGIC & \cite{Colin:2011wc}\\
\hline
\end{tabular}
}
\vspace{-2mm}
\caption{\label{tab:targets}Compound of observational targets for indirect dark
matter searches with gamma rays from the Whipple, H.E.S.S., MAGIC and
VERITAS experiments. In the second and third columns, the year and
duration of observations (in hours) are given. In the last column, link to papers
(or proceedings if papers were not available) are given.}
\end{table}

\subsection{Observation of dwarf satellite galaxies}
\vspace{-1mm}
The Dwarf Satellite Galaxies (DSGs, often called dwarf spheroidal galaxy
or dSphs) are rather small (order $10^7$ M$_\odot$~\cite{Strigari:2008ib}) galaxies,
gravitationally bound to the Milky Way, located in the Milky Way dark
matter halo, at distances below 250~kpc. Cosmological N-body
simulations~\cite{ViaLactea2008,Springel:2008b} predict that, besides the main  
``smooth'' DM halo, a wealth of substructures should be present down
to scales of $10^{-6}-10^{-12}$~M$_\odot$. Onto these overdensities,
baryons could have accreted until starting star and galaxy formation,
however, the impact of baryon physics at these targets is probably smaller
compared to Milky Way sized
objects. Because of the relatively low star content and gravitational pull, most
of the DSG went inactive long time ago, with no major
stellar   activities found. Their total number is an issue, with discrepancies
between observation data and simulations (the so-called ``missing
satellite problem''~\cite{Strigari:2007ma}), and so far only about twenty were
discovered. Also their relation with visible galaxies is subject of debate, because the predictions foresee very large DSG that are not in fact observed, the so-called "too big to fail'' problem~\cite{BoylanKolchin:2011de}. 
They are standardly classified in ``classical'' and
``ultra-faint''. The latters usually refer to those discovered after
the Sloan Digital Sky Survey (SDSS) spectrometric experiment, but
shadows a more subtle difference: ultra-faint DSGs have more compact
sizes, less stars, and are more DM dominated than the  classical, with
mass-to-light ratio sometimes exceeding 1000 $M_\odot/L_\odot$.
Nowadays we know the existence about two dozens DSGs in the MW halo. Among the
classical DSGs we have Sculptor, Fornax, 
Leo I and Leo II, Draco, Ursa Minor, Carina, Sextans while among the ultra-faint DSGs we find: Willman 1, Ursa Major I and II, Hercules, Leo I, II,
IV, Canes Venatici I and II, Coma Berenices, Segue I and Bootes
I and others. DSGs are extremely interesting to observe, for the reason
that the dynamic of the object is strongly steered by DM, and
the baryons play a secondary role. This means that not only the
gamma-ray contribution from possible standard astrophysical sources
(e.g. supernova activity) is low, but also that the determination of the
DM content itself from the stellar velocity distribution is
relatively precisely determinable. The J-factor is computed in DSGs via the
Jeans hydrostatic equilibrium equation from the measured stellar
position and velocities of the DSG member star. To solve the equation
one needs to input a model for the star phase-space 
and another one for the DM density profile, whose normalization is left as a
free parameter. All the details of these calculations are reviewed in
Ref.~\cite{Strigari:2012gn}. As an example, for the Segue~1 DSG,
considered as one of the most dark-matter dominated objects,
Ref.~\cite{Essig:2010a} reports uncertainties in the J-factor of one
order of magnitude: $\log_{10}(J_{\rm
  Segue})=19.0\pm0.6$~GeV$^2$~cm$^{-5}$. 
On the other hand Segue~1 has only 71 confirmed member stars, a factor
ten less than the classical DSGs like Draco, which is probably the
best constrained classical DSGs with a thousand of identified member stars.  

Since 2004, these targets have extensively studied by
IACTs \cite{Wood:2008hx, Albert:2007xg, Acciari:2010ab, Aharonian:2007km, Aharonian:2008dm, Aliu:2008ny, Abramowski:2010aa, Aleksic:2011jx,
  Aliu:2012ga, Aleksic:2013segue}. Sagittarius,
present in Tab.~\ref{tab:targets} is not included 
in the list above of classical DSGs because of its peculiar
nature, with extremely extended tails, which makes very uncertain the estimation of its DM content. However, we mention that this conclusion is under debate, and H.E.S.S. has recently shown results with a total observation time on this target larger than 90~h~\cite{Lamanna:2013yha}.
The table shows that during the first years, different targets have
been observed for limited amount of hours, typically below 20 hours,
while more recently, and specially for the DSG Segue~1, the amount of
hours dedicated to a single source has extremely increased. A first
motivation is that back in 2004, the prospects of detection were more
optimistic~\cite{Bergstrom:2005qk}. At that times, only the classical DSGs were
known, and it was obvious to try to observe one or more of these
objects, and among these the Draco DSG which was considered one of the
most promising. After the advent of the SDSS experiment, and the
continuous discovery of new ultra-faint DSGs, the attention was turned
to these new targets, for whom the predictions were extremely
exciting. 
Obviously, the exclusion limits obtained with the first publications
about the classical DSGs quite improved in the latest
publications, with extended observation times, and often improved
analysis techniques. It is worth mentioning the application of a
dedicated full likelihood method in the deep Segue~1 observation with
the MAGIC-stereo system~\cite{Aleksic:2012cp}, where the authors
improved the sensitivity of the telescopes by up to a factor of 2 by
fully taking into account the specific spectral features of typical DM
spectra, as well as the telescope energy resolution. This method,
applied to DM searches, has so far provided the most constraining
results for indirect DM searches at DSGs with
IACTs, see also Fig.~\ref{fig:compare}. Another specific method based on Event Weighting is developed by VERITAS experiment~\cite{Zitzer:2013cka}, that improves the sensitivity from a joint analysis of individual targets into a single limit and utilizing more of the individual event information.

It is safe to say that the
exclusion limits obtained with DSGs are possibly the most robust for
indirect DM searches with IACTs with upper limits reaching cross-section values of the order of $10^{-24}$~\svu.

\subsection{Observation of globular clusters}
\vspace{-1mm}
The globular clusters share some properties with DSGs, but in general
are less bright and less massive. On the other hand, the stellar
population is normally different, with globular clusters having more
homogeneous stellar content in terms of star classification. The
debate on whether they reside in a DM halo is still
open. From the cosmological point of view, there is no conflict in
telling that they were formed in DM overdensities, as the
DSGs, however, there are no strong observational evidence for the
presence of a DM core. A possibility is also that they
possessed DM cores early disrupted by tidal forces.

The M15, M32 and M33 globular clusters were observed from 2002 to 2004
by Whipple~\cite{Wood:2008hx} and later on M15 was re-observed together
with NGC6388 by H.E.S.S. in 2006--2009~\cite{Abramowski:2011hh}. The best
exclusion curves for annihilating DM come from the H.E.S.S. results, and
are the order of $10^{-24}-10^{-25}$~\svu, however they rely on strong
assumptions of the dominance on DM in these objects.

\subsection{Observation of subhalos}
\vspace{-1mm}We have already observed that standard N-body simulations and
cosmological theories predict the existence of small DM
overdensities at all scales within the main smooth halo. Some of these
``subhalos'' could have been too small to have attracted enough
baryonic matter to initiate star formation and would therefore be
invisible to past and present astronomical observations at all
wavelengths. However, gamma rays could be
expected from those objects due to annihilations or decays of DM~\cite{Pieri:2008a}. A possible
way to observe them is with gamma-ray all-sky monitoring programs,
like those conducted by the EGRET~\cite{Hartman:1999fc} and Fermi-LAT
satellite~\cite{Fermi-LAT:2011iqa}. A population with no counterparts at other
wavelengths and similar spectra could indeed be associated to
subhalos. 

Following this idea, the MAGIC and VERITAS collaboration investigated
among the unidentified Fermi objects (also called UFOs), those with no
obvious counterparts, satisfying some criteria in the stability of the
spectrum, and spectral hardness that could be explained as
subhalos~\cite{Nieto:2011sx}. Few targets were observed by both
experiments, however without any hint of detection, and results are
under publications~\cite{GeringerSameth:2013fra,Nieto:2011uj}. 

\subsection{Observation of intermediate mass black holes}
\vspace{-1mm}There are theories that predict the existence of
black holes of mass comprised between $10^2-10^6$~M$_\odot$, sometimes
referred to as Intermediate Mass Black Holes (IMBHs). Many of them
could reside in the Milky Way halo. They could form as
remnants of collapse of Population III stars (scenario A, $10^2$~M$_\odot$). The average number of
IMBHs in the MW halo was estimated by numerical simulation to be on
the order of about a thousand. In a second case (scenario B,
$10^6$~M$_\odot$), IMBHs originate from massive objects formed directly
during the collapse of primordial gas in early-forming  halos. The
total number of IMBHs in this scenario would be about a hundred.
As a result the increased gravitational potential due to infalling
baryons on a central accreting system, the DM
could have readjusted and shrunk, giving rise to the formation of
what are called ``mini-spikes''. On the other hand, mini-spikes are
rapidly disrupted as 
result of dynamical processes like black hole formation or merging
events. The interesting thing about these objects, is that, starting
from a typical Navarro, Frenk, White (NFW~\cite{Navarro:1997a}) distribution for the DM, the adiabatic
growth of the spike leads to a final DM density profile even
cuspier than the NFW, with central slope of index $-7/3$. 
For scenario B, which foresees more luminous objects than scenario A,
the corresponding gamma-ray luminosity would be of the order of the gamma-ray 
luminosity of the entire Milky Way halo, which made IMBHs very
interesting targets for DM searches~\cite{Bertone:2005xz}. 

As for the subhalo case described above, it was plausible that some
IMBHs could have already been observed during all
sky gamma-ray monitoring programs. After a selection of the EGRET better candidates among
the unidentified sources, MAGIC observed the brightest unidentified
EGRET source in 2006
for 25~h without detection~\cite{Doro:2007icrc}. However, the
source was later on associated with a bright pulsar using Fermi-LAT data. Using 400~h of
data collected from 2004 to 2007 in the region between 
$-30$ and $+60$ degrees in Galactic longitude, and between $-3$ and $+3$
degrees in Galactic latitude, H.E.S.S.~\cite{Aharonian:2008wt} could
exclude scenario B at a 90\% confidence level for dark
matter particles with velocity-weighted annihilation cross-section
\sv~ above $10^{-28}$~\svu~ and mass between 800 GeV and 10 TeV.

\subsection{Observation of galaxy clusters}
\vspace{-1mm}Within the standard $\Lambda$CDM scenario, 
galaxy clusters, with masses around $10^{14}-10^{15}$~M$_\odot$, are
the largest gravitationally bound objects and the most recent structures to
form~\cite{Voit:2004ah}. They are complex objects, relevant for both
cosmological and astrophysical studies, and for what concerns DM
searches \cite[see e.g.][and references therein]{Pinzke:2011ek}. DM, in fact, is supposed to be the
dominant component of the cluster mass budget, accounting for up to 80\%
of its mass (the other components are the galaxies and the gas of the
intra-cluster medium (ICM)). The effect of substructures contribution
to the total DM annihilation flux discussed above is especially effective for
galaxy clusters, where the intrinsic flux ``boost'' from subhalos
can be of order $100-1000$, in particular compared to the case of
DSGs, discussed previously, where the subhalos boost should
contribute only marginally. Despite the fact that, due to their vicinity, DSGs
are usually considered as the best sources for DM indirect detection,
thanks to the subhalos boost, some authors claim that galaxy
clusters have prospects of DM detection better or at least as good as
those of DSGs \cite{SanchezConde:2011ap,Pinzke:2011ek, Gao:2011rf}.
On the other hand, in galaxy clusters, emission in the gamma-ray range
is not only expected by DM annihilation. Clusters may host an Active
Galactic Nucleus (AGN) and radio
galaxies~\cite{Aleksic:Perseus,Aleksic:2010xk}. Moreover gamma-rays
are expected to be 
produced also from the interaction of cosmic rays (CRs) with the
ICM. Such a contribution is usually found to be larger than the one 
predicted from DM annihilation. It thus represents an unavoidable
source of background for DM searches in galaxy clusters. On the other
hand, the different morphologies of DM (extended), cosmic-ray (compact)
and the individual galaxies (point-like) could be used as a
discriminator for the different components, as well as the obvious
differences in the expected gamma-ray spectra from the various
sources~\cite{Doro:2012xx}. 

To date, the deepest exposures are 
performed with the MAGIC stereoscopic system of the Perseus
cluster~\cite{Aleksic:2011cp} and through
the observation of the Coma cluster with VERITAS~\cite{Arlen:2012}. For DM searches, probably the
strongest constraints come from the observation of the Fornax galaxy
clusters, expected to be the most DM dominated one~\cite{Abramowski:2012au}. H.E.S.S., MAGIC and VERITAS have also
made other campaigns on galaxy clusters,
reporting detection of individual galaxies in the cluster, but only
upper limits on any CR and DM associated
emission~\cite{Aleksic:2009ir, Aharonian:2008uq, Aharonian:2009, 
  Acciari:2009uq}.  Even though IACT 
limits are weaker than those obtained from the Fermi-LAT satellite
measurements in the GeV mass range~\citep{Abdo:2010b},
they complement the latter in the TeV mass range.

\subsection{Observation of the galactic center and halo}
\vspace{-1mm}The Galactic Center (GC) is a prime target for DM searches with
IACTs. In 2004, H.E.S.S. published the 
observation of the GC and its interpretation in terms of DM
during an observation campaign of about
50~h~\cite{Aharonian:2006wh}. The resulting gamma-ray spectrum, obtained from a 40
standard deviations significant signal, was consistent
with a power law and no significant features --- as expected at
e.g. the DM mass cutoff --- were detected. The analysis was performed
first considering that the entire gamma-ray flux was coming from dark
matter annihilation (instead of the plausible standard astrophysical
sources like the central BH or closeby SNR). This provides lower
limits on the possible DM mass scale at more than 10~TeV. In
a second analysis, by considering a superposition of a power-law and
different possible DM spectra, the results excluded 
compatibility with any DM model. Later observations from H.E.S.S.
of the galactic center were published in Refs.~\cite{Aharonian:2008yb,
Aharonian:2009zk,Hess:2009tm}, specially focused on a better
localization of the gamma-ray emission, while no additional DM
interpretation was given. Still, questions about the origin of the
gamma rays and the emission mechanism remain open. In short, no conclusive evidence
from DM emission could be drawn from the MW center because of the
strong contribution from background emission of other nature.

For Milky-Way sized DM halo, the radial DM density profiles is obtained
from N-body simulations (e.g. Aquarius [1], Via Lactea II [2]) and
from observations. The former can be described
by Einasto and NFW parameterizations. Large differences between both
parameterizations occur if they are extrapolated down to the very
center of the halo, where the NFW profile is more strongly
peaked. However, at distances larger than 10 pc, the 
difference is reduced to a factor of two. This idea was used by H.E.S.S.
which searched for the DM signal in a region with a projected
galactocentric distance of $45-150$ pc that corresponds to an angular
distance of $0.3-1.0^\circ$. In this ring, in addition, the
contamination from gamma-ray sources is naturally avoided. The results
were presented in Ref.~\cite{Abramowski:2011hc} and resulted in the
most constraining upper limits on DM annihilations from IACTs, as
shown in Fig.~\ref{fig:compare}. We remark that those limits are valid
for cuspy profiles like the NFW and the Einasto ones. However,
experimental observations tends to predict more shallow profile toward
the galactic center (so-called ``cored'' profiles, see, e.g.,
\cite{Donato:2009ab}). All in all, the DM modelization of the
galactic center region remains a complex task, because of the expected
strong interplay with the baryons and the stellar winds and shocks,
still to be clarified (see e.g. discussions in \cite{Gomez-Vargas:2013bea,Ogiya:2013qnf,DiCintio:2013qxa}).

Also MAGIC and VERITAS have observed the Galactic Center in these
years, confirming the results from
H.E.S.S.~\cite{Albert:2005kh,Beilicke:2012rx}. However, this target is
best observable from the Southern Hemisphere, where H.E.S.S. is
located. At the location of MAGIC and VERITAS, the GC culminates low
in the horizon, which results on one hand in a significant higher
energy threshold for observation, but on the other hand in a
considerably higher
sensitivity at high energies than in vertical observation. Both instruments are still conducting
campaigns on this target, and publications are expected in the coming
years. 

\subsection{Constraints for line emission}
\vspace{-1mm}Gamma-ray line signatures can be expected due to self-annihilation or
decay of DM particles in space. Such a signal would be
readily distinguishable from astrophysical gamma-ray sources that in
most cases produce continuous spectra. There have been recent claims
of hint of a line-like signal at about 130~GeV in the Fermi data of
the Galactic Center region~\cite{S_VIB_Line, S_Line} which has
received a huge attention. If confirmed, the WIMP particle should  have a mass
of about $m_\chi\sim130$ GeV  and annihilation rate
(assuming Einasto profile) of  \sv$_{\gamma\gamma} =
1.27\times 10^{-27}$~\svu \cite{Bringmann:2012ez}. 
Using data collected with H.E.S.S., upper
limits on line-like emission were obtained in the energy range between
500 GeV and 25 TeV for the central part of the Milky Way halo and for
extragalactic observations, complementing limits obtained with
the Fermi-LAT instrument at lower energies~\cite{Abdo:2010nc}. No statistically
significant signal could be found. For monochromatic gamma-ray line
emission, flux limits were obtained for
the central part of the Milky Way halo and extragalactic observations,
respectively. For a DM particle mass of 1 TeV, limits on the
velocity-averaged DM annihilation cross section \sv~ reach the level of
$10^{-27}$ \svu, based on the Einasto parameterization of the Galactic DM
halo density profile. Additional limits are calculated in
\cite{Aleksic:2013segue} using the 158~h observation on Segue~1. In
general, they are more than one order of magnitude worse than those
from H.E.S.S. but possibly more solid, because in DSGs a core or a cusp
profile results in J-factors that differ by a factor of few~\cite{Bringmann:2008kj,Charbonnier:2011ft,Martinez:2013els}.

\subsection{Constraints for all-electrons searches}
\vspace{-1mm}Among the cosmic rays impinging the Earth, a small but important
fraction is constituted by electron and positrons (collectively called
``all-electrons'' in this context). When impinging the atmosphere, the
initiate electromagnetic showers, which are indistinguishable from those
initiated by gamma-rays, for what matters IACTs. 

In the last few years, the all-electrons spectrum has shown unexpected
behavior energies at energy larger than 100~GeV. The ATIC balloon
claimed the observation of a bump at above
300~GeV~\cite{Chang:2008a}, which was partially disproved by the very
precise measurements obtained with the Fermi-LAT
satellite~\cite{Abdo:2009a}. Still, in both cases, a discrepancy
between the theoretical predictions and the observational results were
extremely significant. In the same year, H.E.S.S. developed a specific
analysis for electrons, based on the selection of ``electron-like''
events in regions far from gamma-ray sources and subtraction of the
remaining hadronic cosmic ray background using
simulations, and could measure with high precision the
spectrum from 300~GeV to 1.2~TeV using 230~h of observation of the
extragalactic dark sky~\cite{Aharonian:2009a,Aharonian:2008aa}. Later,
MAGIC confirmed these results but with less significance and smaller 
energy coverage~\cite{BorlaTridon:2011dk}. More recently, the AMS-02
satellite provided more precise data (see link \cite{ams-02} that includes data
  from the other experiments mentioned here).
The anomalies in the $e^-$ and $e^+$ fluxes are generally interpreted
as a new component with a harder spectrum and a higher $e^+$ ratio
than what is expected from secondary electrons induced by the 
galactic CR propagation in the galaxy. Many scenarios involving DM
(annihilation/decay), nearby pulsars or inhomogeneity in the 
galactic CRs have been proposed to interpret the data~\cite[see,
  e.g.,][]{Cholis:2008b,Profumo:2008ms}. 

The results obtained with H.E.S.S. and MAGIC are extremely important
in the overall scenario. First of all, they provide an independent
measurement, with different systematics compared to other
experiments. This is specially important because the different results
are sometimes not compatible within the systematics, which could
shadow experimental or intrinsic not yet understood uncertainties. Secondly, the IACTs data could extend to energies larger
than 1~TeV, where AMS-02 and Fermi-LAT are at least less
sensitive. This region is extremely important for the DM searches, in
case of DM mass at the TeV, because it is where a spectral cutoff is
expected. More results are expected on this kind of measurements in
the following years.  

\subsection{Constraints on antiparticles separately}
\vspace{-1mm}In addition to what described in the previous section, in principle,
IACTs can also observe separately electrons and positrons (as well as
proton / antiproton or any charged particle / antiparticle) by using the
fact that if the telescope is observing at a certain distance close
enough to the Moon, the Moon itself blocks some cosmic rays creating a
hole in the flux which is shifted 
along an axis perpendicular to the geomagnetic field and verse
depending on the charge of the particle. The amplitude
and direction of the deviation depend on the energy and charge of the
cosmic ray. For example, at 500\,GeV, the electron or positron shadow is shifted
3-4$^\circ$ away from the Moon. This technique is currently under
investigation by the MAGIC~\cite{Colin:2011wc} and VERITAS
experiment. However, no results 
have been published yet, and the feasibility of the technique has still
to be proved. 

The shape of the $e^+$ ratio above 100\,GeV, particularly at the
$300-800$\,GeV range, is crucial to discriminate 
between these models and to understand the origin of the electron
spectrum anomalies. PAMELA, Fermi-LAT and AMS-02 among others,
measured the $e^+$ ratio up to about
350\,GeV~\cite{ams-02-positron}. Again, ground-based telescopes
have in principle sensitivities larger than satellite for positrons
above 350~GeV. Currently, the positron spectrum has very smooth
features from 10 to 350~GeV, and therefore, the observation of some
spectral features above this range would constitute an extremely
important probe for possible DM interpretations.

\section{The future with the Cherenkov Telescope Array}
\label{sec:cta}
\vspace{-3mm}
The Cherenkov Telescope Array (CTA)~(see Ref.~\cite{CTAConsortium:2010a}
for a full description and Ref.~\cite{Doro:2009qs} for a shorter version
also presented at this conference series) is a
project for a next-generation 
observatory for very high energy (GeV--TeV) ground-based gamma-ray
astronomy, currently in its design phase, and foreseen to be operative
a few years from now. Several tens of telescopes of 3--4 different
sizes, distributed over a large area, will allow for a sensitivity
about a factor 10 better than current instruments such as H.E.S.S, MAGIC and
VERITAS, an energy coverage from a few tens of GeV to several tens of
TeV, and a field of view of up to 10~deg. The search for new physics
beyond the Standard Model (SM) of particle physics is among the key
science drivers of CTA~\cite{Hinton:2013}. 

CTA is expected to improve the prospects for DM searches with
IACTs on the following basis: $a)$ naively, the improved sensitivity, compared to 
  current instruments, will generically improve the probability of
  detection, or even \emph{identification} of DM, through the
  observation of spectral features, $b)$ the energy range will be
extended. At low energies, this will allow overlap with 
  the Fermi-LAT or other similar future instruments, and will provide sensitivity to WIMPs with
  low masses. For WIMPs with mass larger than about 100 GeV, CTA will
  be the experiment with highest sensitivity;  $c)$ the increased FOV (about 10 deg versus $2-5$ deg)
  with a much more homogeneous sensitivity, as well as the improved
  angular resolution, will allow for much more efficient searches for
  extended sources like galaxy clusters and
  spatial anisotropies; $d)$ finally, the improved energy resolution
  will allow much better 
  sensitivity to the possible spectral feature in the DM-generated
  photon spectrum. The observation of a few identical
  such spectra from different sources will allow both precision
  determination of the mass of the WIMP and its annihilation
  cross-section.

In Ref.~\cite{Doro:2012xx}, a detailed report of the performance for
CTA DM searches, estimated employing Monte Carlo simulations of different possible
CTA realizations, is given, as well as for other fundamental
physics issues, like the possible existence of axion-like particles, expected
violation of Lorentz Invariance by quantum gravity effects. 
In \cite[Fig.~23]{Doro:2012xx}, the expected performance of CTA for
WIMPs annihilating purely into $b\bar{b}$ in 100~h observation at
DSGs, galaxy clusters, and at the Galactic Center Halo is shown,
together with extrapolation of the Fermi-LAT performance for 10~years
of data. As expected, the best results
are achieved for the observation in the vicinity of the Galactic
Center, where the  thermal annihilation cross-section for
WIMP DM of $10^{-26}\,\mbox{cm}^3\mbox{s}^{-1}$ is at reach. This
would be the first time that ground-based Cherenkov telescopes could
reach this sensitivity level.  

It is worth noting here that on July 2012 the H.E.S.S. collaboration
inaugurated the fifth telescope, a 28~m diameter dish telescope which
is currently the largest Cherenkov dish ever
built~\cite{Giebels:2013dxa}. This will result in a larger sensitivity
below 100 GeV with interesting expectations 
specially for the 130~GeV line searches~\cite{Bergstrom:2012vd}.
 
\section{Conclusions}
\label{sec:conclusions}
\vspace{-3mm}
\begin{figure}[t]
  \centering
  \includegraphics[width=0.95\linewidth]{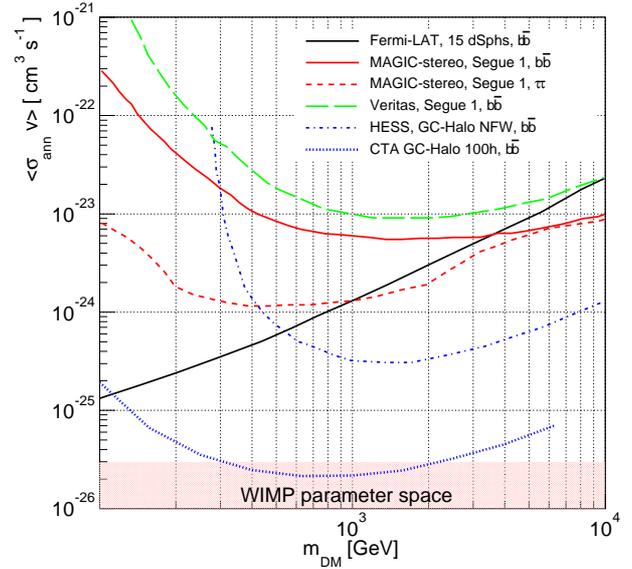}
\vspace{-3mm}
  \caption{\label{fig:compare} Comparison of some exclusion lines
    for the Fermi-LAT observation of 15 combined DSGs for $b\bar{b}$
    (solid black)~\cite{Ackermann:2013yva},
    H.E.S.S. observation of the galactic center halo for the NFW
    (dot-dashed blue~\cite{Abramowski:2011hc}) for the $b\bar{b}$
    channel, MAGIC-stereo observations of the Segue~1 DSG for the 
    $b\bar{b}$ (solid red) and $\tau^+\tau^-$ (dashed red)
    channels~\cite{Aleksic:2013segue}, Veritas observations of the
    Segue~1 DSG for $b\bar{b}$ (dashed green) \cite{Aliu:2012ga},  
    and for the estimation for 100~h observation at the galactic
    center halo with CTA (thick dashed blue)~\cite{Doro:2012xx}. More
    details in the text.} 
\end{figure}
%
In Fig.~\ref{fig:compare} we collect few results on the
exclusion curves for WIMP annihilation cross-section. On the bottom
left side of the plot, we see the exclusion 
power of Fermi-LAT observation of 15 combined DSGs for the $b\bar{b}$
(solid black line)~\cite{Ackermann:2013yva}
We also show the best limit obtained on
DSG observation with Segue~1 observed during 158~h with the MAGIC
stereo experiment again for the  $b\bar{b}$ (solid red line) and
$\tau^+\tau^-$ (dashed red line) channels~\cite{Aleksic:2013segue}.
These two channels somehow represent
two extreme cases, a very soft spectrum ($b\bar{b}$) and a very hard
spectrum ($\tau^+\tau^-$). 
One can see that because of the better sensitivity of MAGIC at higher
energies, the harder $\tau^+\tau^-$ channel is better constrained. The
same target was observed in 48~h of observation with Veritas (dashed green)
\cite{Aliu:2012ga}. 
In blue, we show the
H.E.S.S. exclusion curve from the galactic center halo for the NFW
(dot-dashed blue line) \cite{Abramowski:2011hc} for the $b\bar{b}$ channel only.
Finally, we show estimates for 100~h observation of the
Galactic Center halo region with CTA (thick dashed blue,
\cite{Doro:2012xx}) considering again a NFW profile.

The detection of gamma-rays provides complementary
information to other experimental probes of particle DM, especially
that of direct detection, because CTA could be able to access a
fraction of the parameter space not accessible
otherwise~\cite{Bergstrom:2010gh}. With respect to
particle searches at the LHC, the comparison is not straightforward,
as LHC results are usually strongly related to specific models, and
general conclusions are somewhat model dependent.
%
In any case, LHC discovery of DM,
would prompt the need for proof that the particle is actually consistent with the
astrophysical DM. A concrete scenario has been
analyzed~\cite{Bertone:2011pq} in the case of a SUSY model in the
so-called 
co-annihilation region. Simulated LHC data were used to derive
constraints on the particle physics nature of the DM, with the result
that the LHC alone is not able to reconstruct the neutralino
composition. The situation improves if the information from a
detection of gamma-rays after the observation of the Draco DSG by
IACT like CTA is added to the game: the internal
degeneracies of the 
SUSY parameter space are broken and including IACT results allows us to fully
interpret the particle detected at the LHC as the cosmological DM.  
In 
the other case where the LHC will not detect any physics beyond the
Standard Model, predictions were
made in the context of the CMSSM~\cite{Bertone:2011kb} indicating that
the mass of the neutralino will be bound to be close to the TeV scale.
In this scenario, MAGIC, H.E.S.S. and VERITAS and,
even more, CTA could be the only
instrument to be able to detect and identify a WIMP candidate with
masses beyond some hundreds GeV. Finally, we mention that CTA will
open a new possibility in detecting DM with IACTs based on
the detection of spatial anisotropies in the diffuse extragalactic
gamma-ray sky~\cite{Fornasa:2012gu,Ripken:2012db}.

\vspace*{3mm}
\footnotesize{{\bf Acknowledgment:}{ I gratefully acknowledge
    J.~Conrad, C.~Farnier, R.~Ong, M.~Fornasa and A.~Smith for
    comments on the manuscript as well as the MAGIC and CTA
    collaborations. This work is funded by University of Padova.} 


\bibliographystyle{model1-num-names}
\vspace{-3mm}
\bibliography{biblio}

\end{document}